\documentclass[preprint2]{aastex}

\shorttitle{The birth rate of SNe Ia from hybrid CONe WDs}
\shortauthors{Meng \& Podsiadlowski}

\begin{document}


\title{The birth rate of SNe Ia from hybrid CONe white dwarfs}


\author{Xiangcun. Meng}
\affil{$^1$ National Astronomical Observatories/Yunnan
Observatory, the Chinese Academy of Sciences, Kunming, 650011,
China, \\xiangcunmeng@ynao.ac.cn\\
$^2$ Key Laboratory for the Structure and Evolution of Celestial
Objects, Chinese Academy of Sciences, Kunming 650011, China\\}
\and
\author{Philipp. Podsiadlowski}
\affil{$^3$ Department of Astronomy, Oxford University, Oxford OX1 3RH} %\email{aastex-help@aas.org}

%\and

%\author{R. J. Hanisch\altaffilmark{5}}
%\affil{Space Telescope Science Institute, Baltimore, MD 21218}


%\altaffiltext{1}{Visiting Astronomer, Cerro Tololo Inter-American Observatory.
%CTIO is operated by AURA, Inc.\ under contract to the National Science
%\altaffiltext{2}{Society of Fellows, Harvard University.}
%\altaffiltext{3}{present address: Center for Astrophysics,
%    60 Garden Street, Cambridge, MA 02138}
%\altaffiltext{4}{Visiting Programmer, Space Telescope Science Institute}
%\altaffiltext{5}{Patron, Alonso's Bar and Grill}


\begin{abstract}
Considering the uncertainties of the C-burning rate (CBR) and the
treatment of convective boundaries, \citet{CHENM14} found that there
is a regime where it is possible to form hybrid CONe white dwarfs
(WDs), i.e.\ ONe WDs with carbon-rich cores. As these hybrid WDs can
be as massive as 1.30\,$M_{\odot}$, not much mass needs to be accreted
for these objects to reach the Chandrasekhar limit and to explode as
Type Ia supernovae (SNe Ia). We have investigated their contribution
to the overall SN Ia birth rate and found that such SNe Ia tend to be
relatively young with typical time delays between 0.1 and 1 Gyr, where
some may be as young as 30\,Myr. SNe Ia from hybrid CONe WDs may
contribute several percent to all SNe Ia, depending on the
common-envelope ejection efficiency and the CBR. We suggest that these
SNe Ia may produce part of the 2002cx-like SN Ia class.

\end{abstract}





\section{INTRODUCTION}\label{sect:1}
As very good cosmological distance indicators, Type Ia supernovae
(SNe Ia) have been successfully used for determining basic
cosmological parameters; this has led to the discovery of an
accelerating expansion of the Universe (\citealt{RIE98};
\citealt{PER99}). However, the exact nature of SNe Ia is still not
clear, especially concerning their progenitor systems
(\citealt{HN00}; \citealt{LEI00}); indeed, the identification of
their progenitors is important for several astrophysical fields
(\citealt{WANGB12}). It is for about four decades that two basic
scenarios for the progenitor of SNe Ia have been competing. In the
single-degenerate (SD) model, a carbon-oxygen white dwarf (CO WD)
grows in mass via accretion from a non-degenerate companion
(\citealt{WI73}; \citealt{NTY84}) while, in the double-degenerate
(DD) scenario, two WDs merge after losing orbital angular momentum
by gravitational wave radiation (\citealt{IT84}; \citealt{WEB84}).
At present, there is some observational and/or theoretical support
for both basic scenarios, but there are also counter-arguments
(e.g., \citealt{MAOZ14}; \citealt{RUIZLAPUENTE04}).

The discovery of circumstellar material from the SN Ia progenitor
system has provided significant evidence in support of the SD
scenario (\citealt{PAT07}; \citealt{STERNBERG11};
\citealt{DILDAY12}), but a major shortcoming of the model is that
present estimates of their birth rate appear to be somewhat lower
than the observationally inferred rate (\citealt{HAN04};
\citealt{MENGYANG10}; also see \citealt{HAC99a}; \citealt{YUN00};
\citealt{RUITER09}; \citealt{Mennekens10}; \citealt{CLAEYS14}).
One reason for the low birth rate is that it is difficult to
increase the WD mass to reach the Chandrasekhar limit
(\citealt{BRANCH95}; \citealt{HOWEL11}; \citealt{MAOZ12};
\citealt{MAOZ14}; ). Recently, following the discovery of
\citet{DENISSENKOV13}, \citet{CHENM14} found that considering the
uncertainty of the C-burning rate (CBR) and the treatment of
convective boundaries, hybrid CONe WDs with a carbon-rich core may
form instead of pure ONe WDs even for stars with an initial mass
larger than 7.0\,$M_{\odot}$. In their most extreme case (for a
CBR efficiency factor of 0.1), the hybrid WD could be as large as
1.30\,$M_{\odot}$. It is relatively easy for such WDs to accrete
enough mass to reach the Chandrasekhar limit, i.e.\ the discovery
by \citet{CHENM14} could increase the fraction of stars that form
WDs capable of igniting carbon in a thermonuclear runaway and
contribute to the birth rate of SNe Ia. \citet{CHENM14} did not
yet estimate the birth rate of SNe Ia from hybrid CONe WDs. The
purpose of this paper is to provide such estimates using binary
population synthesis (BPS) based on the results in
\citet{CHENM14}.

In Section \ref{sect:2}, we describe our method and present the
results of our calculations in Section \ref{sect:3}. In Section
\ref{sect:4}, we conclude with a short discussion of the implications.

\begin{figure*}
\centerline{\includegraphics[angle=270,scale=.7]{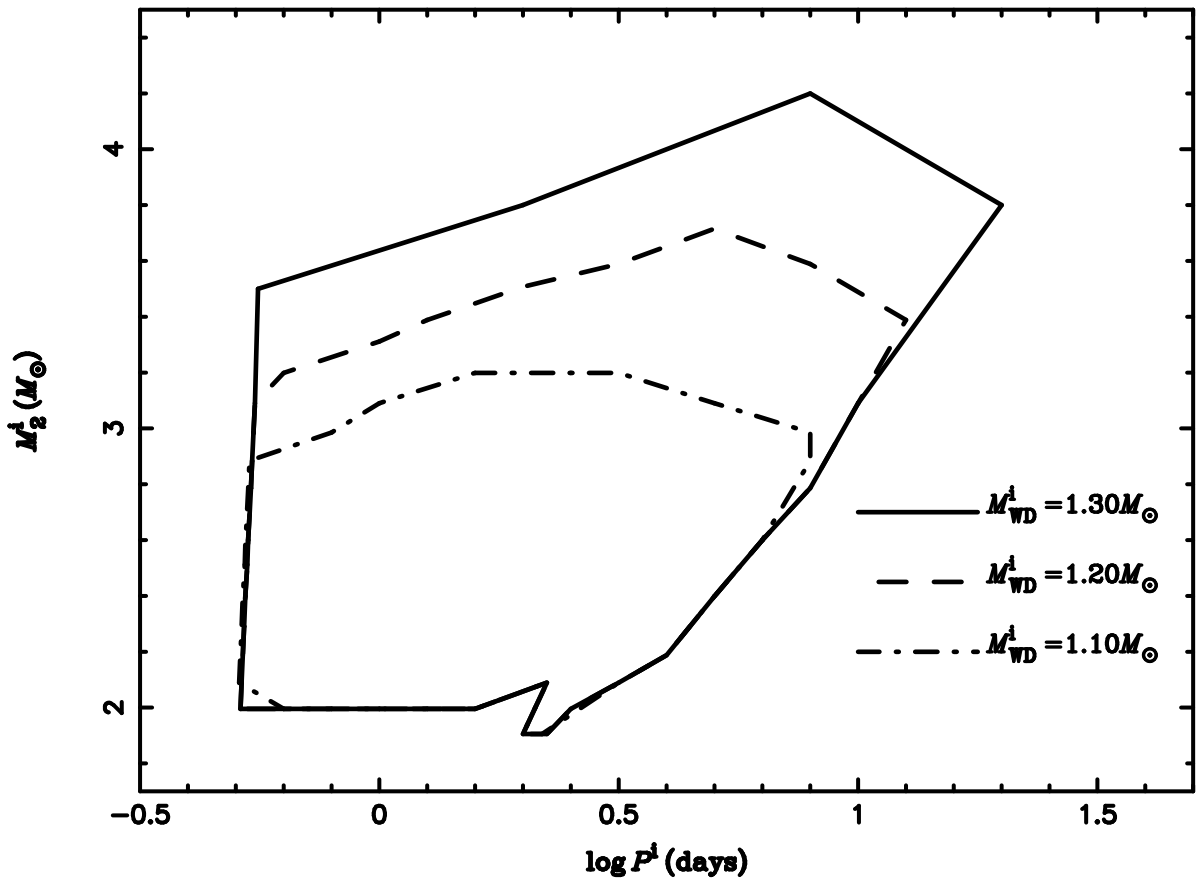}}
\caption{Contours in the orbital period -- secondary mass plane for WD
  binaries leading to SNe Ia for different initial WD masses (as
  indicated).}\label{cour}
\end{figure*}

\section{METHOD}
\label{sect:2}
%% In a manner similar to \objectname authors can provide links to dataset
%% hosted at participating data centers via the \dataset{} command.  The
%% second curly bracket argument is printed in the text while the first
%% parentheses argument serves as the valid data set identifier.  Large
%% lists of data set are best provided in a table (see Table 3 for an example).
%% Valid data set identifiers should be obtained from the data center that
%% is currently hosting the data.

%% Note that AASTeX interprets everything between the curly braces in the
%% macro as regular text, so any special characters, e.g. "#" or "_," must be
%% preceded by a backslash. Otherwise, you will get a LaTeX error when you
%% compile your manuscript.  Special characters do not
%% need to be escaped in the optional, square-bracket argument.
To estimate the birth rate of SNe Ia from hybrid CONe WDs, we do
not new binary evolution calculations but can use previously
published ones.  Based on the optically thick wind model
(\citealt{HAC96}), \citet{MENG09} already obtained a dense model
grid leading to SNe Ia with different metallicities and initial WD
masses. In their calculations, they only considered the case of
main sequence or sub-giant companions (WD + MS). Here, we also
only consider the WD + MS case since the contribution to the total
SNe Ia from WD binaries with red-giant companions is quite
uncertain (e.g. \citealt{YUN95}; \citealt{HAC99b};
\citealt{HAN04}). Using the results of \citet{MENG09}, we simply
extrapolate the WD mass by a linear assumption to obtain the
parameter space leading to SNe Ia for $M_{\rm WD}^{\rm i}=1.30$
$M_{\odot}$. Fig. \ref{cour} shows the contours leading to SNe Ia
for different initial WD masses.

To obtain the birth rate from hybrid CONe WDs, we carried out a
series of detailed Monte Carlo simulations with the rapid binary
evolution code developed by \citet{HUR00,HUR02}. We assumed that,
if a WD is less massive than the most massive hybrid one shown in
Fig.\ 5 of \citet{CHENM14} and is not a CO WD, it is a hybrid CONe
WD. If a binary system in the simulations evolves to the CONe WD +
MS stage and the system is located in the ($\log P^{\rm i}$ --
$M_{\rm 2}^{\rm i}$) plane for a SN Ia at the onset of Roche-lobe
overflow (RLOF), we assume that carbon may be ignited in the
center of the WD no matter how massive the CO core is in the
hybrid WD, and then a SN Ia is produced. We follow the evolution
of $10^{\rm
  8}$ sample binaries. This evolutionary channel is described in more
detail in \citet{MENG09}. As in \citet{MENG09}, we adopted the
following input for the simulations: (1) A single starburst (where
$10^{\rm 11} M_{\odot}$ of stars are formed at the same instant of
time) or a constant star formation rate over the last 15 Gyr. (2)
The initial mass function (IMF) of \citet{MS79}. (3) A constant
mass-ratio distribution is taken to be constant. (4) The
distribution of separations in $\log a$ for wide binaries, where
$a$ is the orbital separation. (5) Circular orbit for all
binaries. (6) The common envelope (CE) ejection efficiency
$\alpha_{\rm CE}$, which denotes the fraction of the released
orbital energy used to eject the CE, is set to 0.5, 0.75, 1.0 or
3.0. (See \citet{MENG09} for details). In this paper, we do not
test the effect of other inputs to produce the binary samples on
the final results such as different IMF, since they may not change
the basic conclusion significantly (see also \citealt{WANGB13}).

%% In this section, we use  the \subsection command to set off
%% a subsection.  \footnote is used to insert a footnote to the text.


%% This section also includes several of the displayed math environments
%% mentioned in the Author Guide.

\section{RESULTS}\label{sect:3}

\begin{figure*}
\centerline{\includegraphics[angle=270,scale=.7]{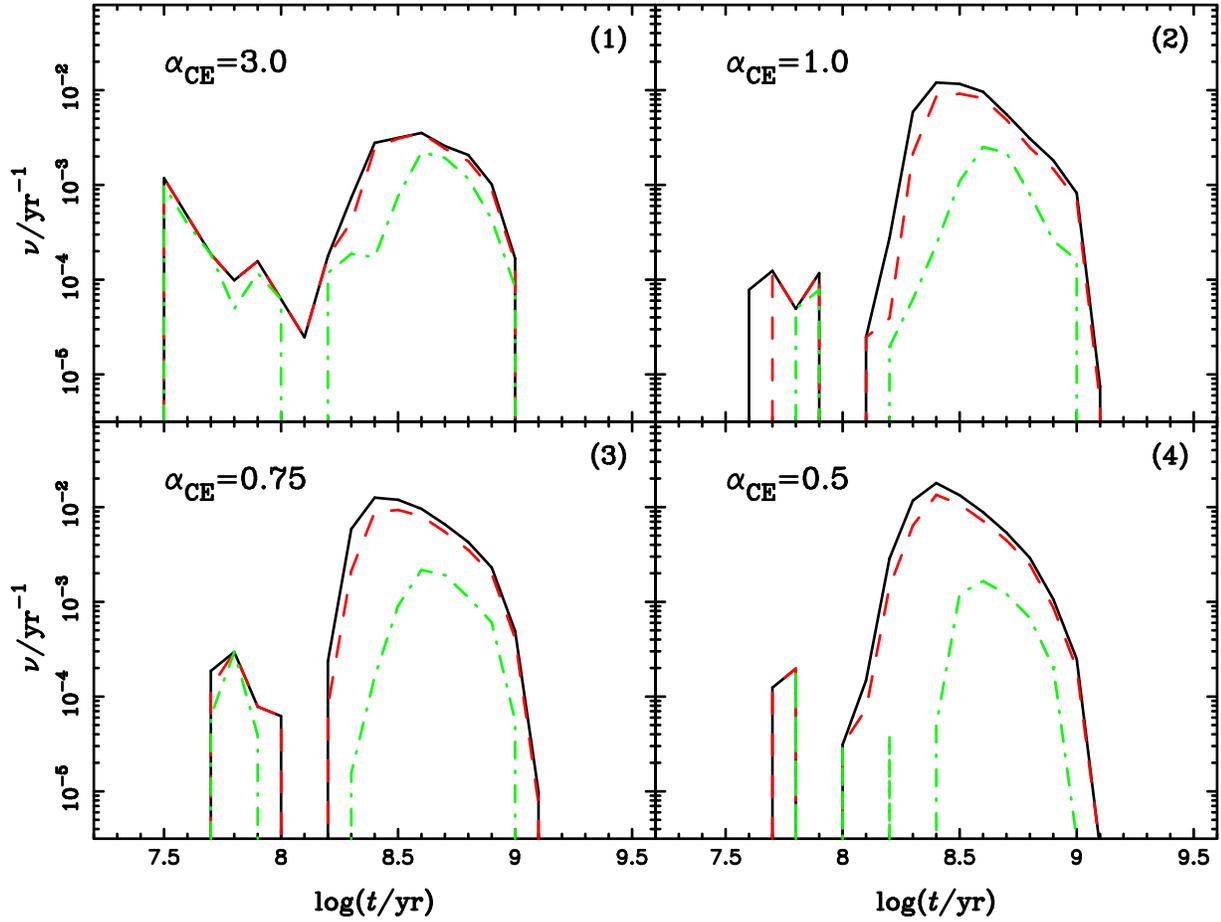}}
\caption{The evolution of the birth rate of SNe Ia from hybrid CONe
  WDs for a single starburst of $10^{\rm 11}M_{\odot}$ for different
  values of $\alpha_{\rm CE}$ (as indicated in each panel). The solid,
  dashed and dot-dashed curves represent the cases for CBR factors of
  0.1, 1 and 10 based on the Fig. 5 in \citet{CHENM14},
  respectively.}\label{single}
\end{figure*}

\begin{figure*}
\centerline{\includegraphics[angle=270,scale=.7]{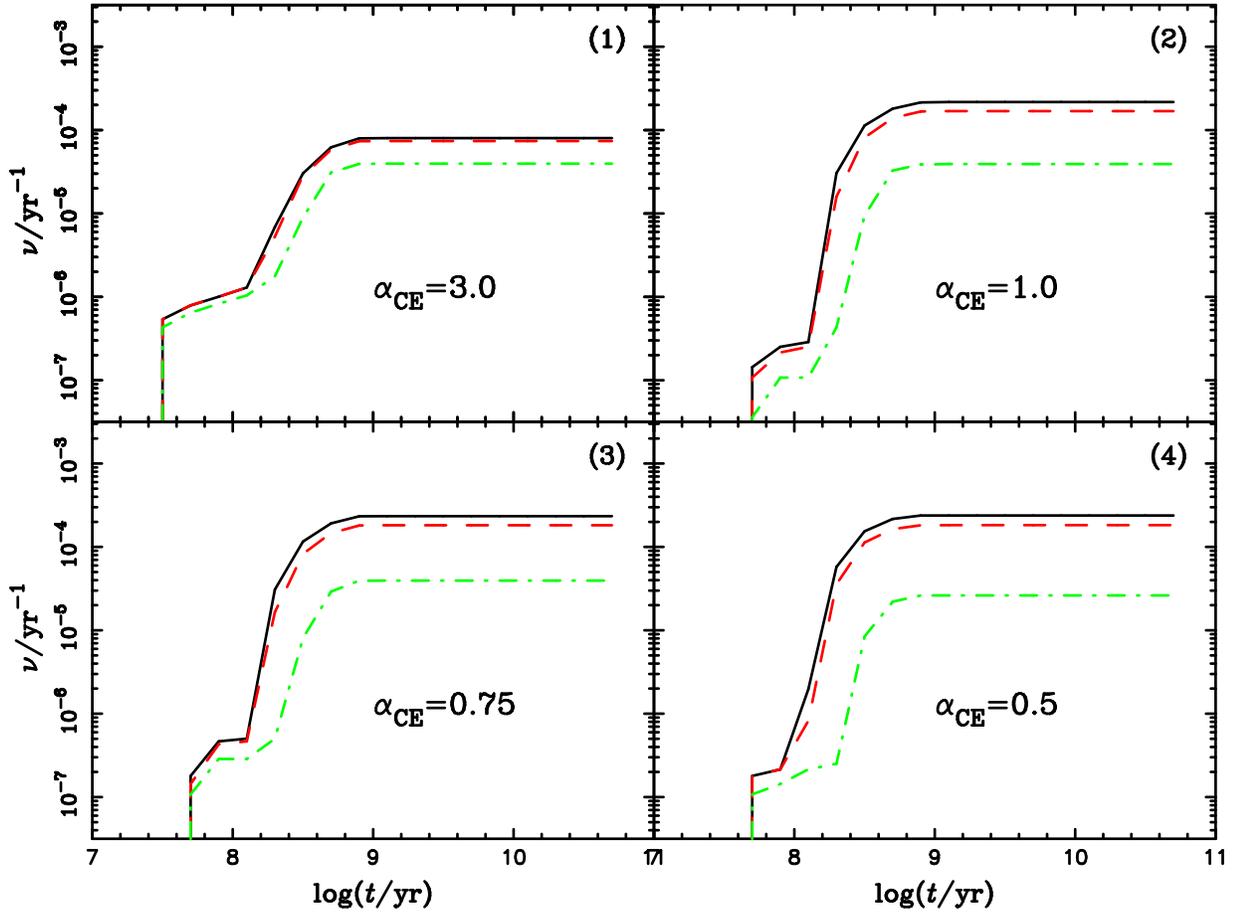}}
\caption{The evolution of the birth rates of SNe Ia from the hybrid
  CONe WDs for a constant star formation rate for different values of
  $\alpha_{\rm CE}$ (as indicated in each panel). The solid, dashed
  and dot-dashed curves represent the cases for CBR factors of 0.1, 1
  and 10 based on the Fig. 5 in \citet{CHENM14},
  respectively.}\label{sfr}
\end{figure*}

The birth rate of SNe Ia for a single starburst from our BPS
simulations is presented in Fig.~\ref{single}. It shows that most
supernovae occur between 0.1 and 1\,Gyr after a starburst, even though
some SNe Ia can be as young as 30\,Myr. The contribution of these
extremely young SNe Ia decreases with decreasing $\alpha_{\rm
  CE}$. These extremely young SNe Ia come from the He star channel, as
defined in \citet{MENG09}, where the first mass-transfer phase in
the primordial binary occurs when the original primary crosses the
Hertzprung gap (HG) or is on the red-giant branch (RGB). In this
case, mass transfer leads to the formation of a common envelope,
and the primary becomes a He star after its ejection. The helium
star fills its Roche lobe again after central helium is exhausted
(so-called case BB mass transfer). Since the mass donor is much
less massive than before, the second phase of RLOF is dynamically
stable, resulting in a close WD+MS system where the companion is
helium-rich. In this binary channel, even a star as massive as
$8-10\,M_{\odot}$ can avoid the fate of a core-collapse supernova
and form a WD. However, this channel is different from the one
described in \cite{CHENM14}, and it is still unclear whether the
WD from such channel is a hybrid CONe WD or not. Irrespectively,
SNe Ia from this channel are rare (see also Fig. \ref{sfr}).

Fig.~\ref{single} also shows that a low $\alpha_{\rm CE}$ leads to a
higher birth rate (see also Fig.~\ref{sfr}) since, for a low
$\alpha_{\rm CE}$, the primordial system needs to release more orbital
energy to eject the CE to form a WD + MS system; this produces WD + MS
systems that tend to have shorter orbital periods which more easily
fulfill the condition for SNe Ia. At the same time, more and more
systems which could otherwise pass through the He star channel merge
with decreasing $\alpha_{\rm CE}$; this results in a decrease of the
number of SNe Ia from this channel. Fig.~\ref{single} also shows that
the birth rate decreases with the CBR factor, but the difference
between a CBR factor of 0.1 and 1 is not significant as the very
massive WDs are rare.

Fig.~\ref{sfr} shows the Galactic birth rates of SNe Ia for a
constant star formation rate (${\rm SFR} = 5.0\,M_{\odot}/{\rm
yr}$) from hybrid CONe WD. The Galactic birth rate is around
$0.26-2.4 \times10^{\rm -4}{\rm yr^{\rm -1}}$, depending on
$\alpha_{\rm CE}$ and the CRB factor. This is much lower than that
inferred overall SN Ia rate from observations ($3-4 \times10^{\rm
-3}{\rm yr^{\rm -1}}$, \citealt{VAN91}; \citealt{CT97}). Hence
these can only contribute between 0.65\,\% and 8\,\% of the total
SN Ia rate. Again, the Galactic birth rate increases in line with
an decreasing CBR factor and $\alpha_{\rm CE}$.

\begin{figure*}
\centerline{\includegraphics[angle=270,scale=.7]{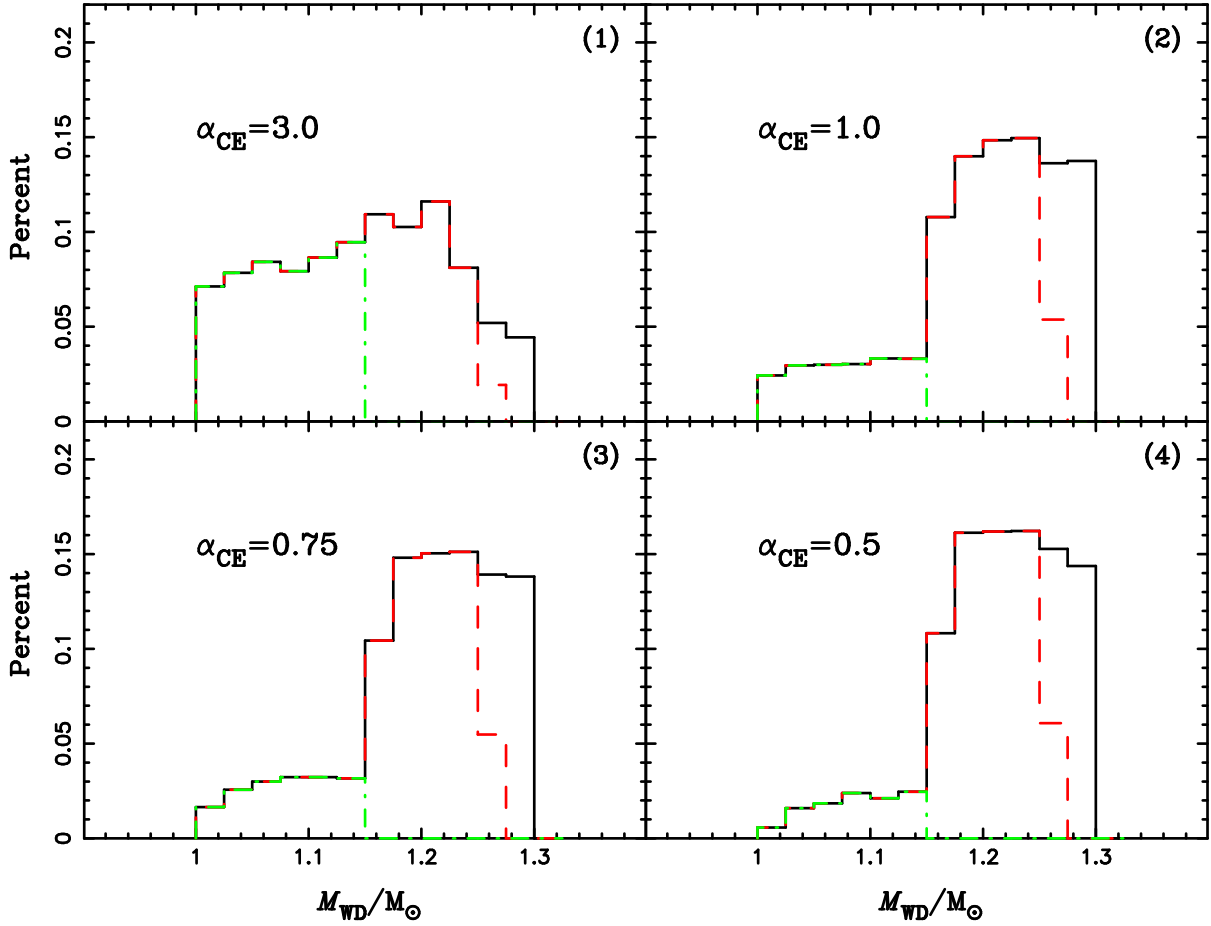}}
\caption{The distribution of the initial hybrid CONe WD masses for
different values of $\alpha_{\rm CE}$. The solid, dashed and
dot-dashed histograms represent the cases for CBR factors of 0.1,
1 and 10 based on the Fig.~5 in \citet{CHENM14},
respectively.}\label{mwddis}
\end{figure*}

Fig.~\ref{mwddis} shows the mass distribution of the initial
masses of the hybrid CONe WDs. Most of the WDs are initially more
massive than 1.05$\,M_{\odot}$ (the upper limit for CO WDs in
Fig.~5 of \citet{CHENM14} for a CBR factor of 1, see also
\citet{MENG08}); i.e.\ irrespective of the correct value of
$\alpha_{\rm CE}$, most SNe Ia come from the channel described by
\citet{CHENM14}, and the contribution from the He star channel is
only minor. There is a relatively small difference of distribution
between CBR factors of 0.1 and 1, as well as the birth rate as
shown in Figs.~\ref{single} and \ref{sfr}.

 \section{DISCUSSIONS AND CONCLUSIONS}\label{sect:4}
\subsection{Uncertainties}\label{sect:4.1}

In this paper, we examined the evolution of the birth rate of SNe
Ia from hybrid CONe WDs, proposed by \citet{CHENM14}, and found
that such SNe Ia could potentially contribute between roughly 1
and 8\,\% of the overall SN Ia rate. The two main uncertainties
are the CBR factor and $\alpha_{\rm CE}$. All of these estimates
are based on an assumption that, if a WD is less massive than the
most massive hybrid one shown in Fig.~5 of \citet{CHENM14} and is
not CO WD, it is a hybrid CONe WD; however, the boundary between
the CO WD and the hybrid CONe WDs here is based on a CBR factor of
1 in the Hurley code, and even for the CBR factor of 1, the mass
limit of initial main sequence stars for carbon ignition in
\citet{CHENM14} is slightly higher than that in the Hurley code
for a relatively smaller convective overshooting in
\citet{CHENM14}. So it is possible that, for a CBR factor of 0.1
and 1, we could overestimate the birth rate of SNe Ia while, for a
CBR factor of 10, we could underestimate it. The results presented
here also include SNe Ia from the He star channel, which is
different from the suggestion by \citet{CHENM14}. At present, it
is unclear whether this channel can produce hybrid CONe WDs.
Fortunately, the contribution for hybrid CONe WDs from the He star
channel is small, i.e. those with an initial mass less than
$1.05\,M_{\odot}$ (see Figs~\ref{sfr} and \ref{mwddis}); therefore
the effect of these uncertainties on our final results is not
significant. In addition, we only considered the case of WD+MS
channel, but a WD of 1.30\,$M_{\odot}$ could also reach the
Chandrasekhar limit by wind or normal Roche lobe overflow for a RG
donor. At present, the fraction of SNe Ia from the RG channel is
very uncertain, but believed to be small (\citealt{YUN95};
\citealt{HAN04}; \citealt{RUITER09}; \citealt{MENGYANG10};
\citealt{WANGB10}; \citealt{Mennekens10}); and therefore this
channel probably does not significantly add to our estimates of
the SN Ia rate with CONe WDs. Moreover, although we assumed that
all the hybrid WDs may produce SNe Ia, it is indeed unclear what
is the smallest C-core mass to make thermonuclear ignition at
present. If there is such smallest C-core mass, the birth rate of
SNe Ia from the hybrid CONe WDs in this paper should be taken as
an upper limit. Furthermore, CE is very important for the
formation of WD + MS system (see \citealt{MENG09}), while whether
a CE forms or not depends on the comparison of a donor star's
radial response to mass loss with the response of its Roche
radius. Recently, the response of fully convective giants to rapid
mass loss has been severely questioned (\citealt{WOODS11};
\citealt{PASSY12}), which means that it becomes relatively
difficult to form a CE, and then relatively difficult for a WD +
MS system to fulfill the condition leading to SNe Ia. So,
according to the discussions above, we conclude that a
conservative upper limit for the contribution to all SNe Ia from
the hybrid CONe WDs is about 10\,\%.

\subsection{The effect of the CBR factor}\label{sect:4.2}
One of the motivations in \citet{CHENM14} comes from the
uncertainty of the carbon burning rate. Here, we only explored
three values for the CBR factor. However, based on the WD mass
distribution in Fig.~\ref{mwddis} and the relation between hybrid
WD mass and the CRB factor in Fig.~5 of \citet{CHENM14}, if the
CRB factor is larger than $\sim 400$, hybrid CONe WDs could
\textbf{not} contribute to the SN Ia rate. Whatever, the
contribution to all SNe Ia from the hybrid CONe WDs decreases with
increasing CRB factor. In addition, from Fig.~5 of
\citet{CHENM14}, one may expect that if the CRB factor were as
large as 100 or 1000, the birth rate of SNe Ia from the SD model
should be much smaller than the present estimations.

\subsection{The properties of SNe Ia from  hybrid CONe WDs}\label{sect:4.3}
The present study shows that SNe Ia from hybrid CONe WDs are
relatively young and could be as young as 30\,Myr. Such SNe Ia may
follow the star formation in late-type galaxies. In addition, compared
with normal CO WDs, hybrid CONe WDs have a relatively low carbon
abundance. If the maximum luminosity of SNe Ia is determined by the
carbon abundance, i.e.\ a low carbon abundance leads to a dimmer SN Ia
(\citealt{NOM03}), SNe Ia from hybrid CONe WDs should have a lower
peak luminosity.  Moreover, for the same reason, a low explosion
energy could be expected, i.e.\ such SNe Ia have a relatively low
kinetic energy per unit mass.  Finally, for SNe Ia from the He star
channel, the accreted material by the hybrid CONe WDs is helium-rich,
which could lead to the detection of helium lines in early spectra of
such SNe Ia.

\subsection{A possible progenitor for 2002cx-like supernova?}\label{sect:4.4}
2002cx-like SNe Ia (referred to as Type Iax supernovae by
\citealt{FOLEY13}) are excellent candidates for observational
counterparts of SNe Ia from CONe WDs. They exhibit iron-rich
spectra at early phases like SN 1991T, while the luminosity may be
as low as that of the faint SN 1991bg and the expansion velocity
is roughly half of those of normal SNe Ia (\citealt{LIWD03}). A
few such events show helium lines in their spectra
(\citealt{FOLEY13}). Furthermore, 2002cx-like SNe Ia favour
late-type galaxies.  Their contribution to the overall SN Ia rate
is quite uncertain due to the heterogeneity of this subclass
(\citealt{NARAYAN11}): estimates of their fractional contribution
range from $5.7^{\rm +5.5}_{\rm -3.8}$ (\citealt{LIWD11}) to
$31^{\rm +17}_{\rm -13}$\,\% (\citealt{FOLEY13}). One of the main
causes for these differences arises from the uncertainty whether
very sub-luminous SNe like SN 2008ha  should be included in the
group or not. The above properties of 2002cx-like SNe Ia are quite
similar to those from hybrid CONe WDs. Considering the uncertainty
of the fraction for SN 2002cx-like objects and taking into account
that at least one 2002cx-like event (SN~2008ge) is hosted by a S0
galaxy with no signs of star formation, we suggest that SNe Ia
from the hybrid CONe WDs might explain part of the SN 2002cx-like
population. Another part could be from double detonation
explosions where a CO WD accretes helium-rich material from a
helium star (\citealt{WANGB13}) and some could actually be due to
fall back in a core-collapse supernova (\citealt{MORIYA10}).
Irrespective of these uncertainties, we encourage numerical
simulations of thermonuclear explosions of hybrid CONe WDs to
further explore our suggestion.

\section*{Acknowledgments}
We are grateful to the anonymous referee for his/her constructive
comments. This work was partly supported by NSFC (11273012,
11033008) and the Key Laboratory for the Structure and Evolution
of Celestial Objects, Chinese Academy of Sciences.

\end{document}